\documentclass[lettersize,journal]{IEEEtran}

\ifCLASSINFOpdf
   \usepackage[pdftex]{graphicx}
   \graphicspath{{figures/}}
   \DeclareGraphicsExtensions{.pdf,.jpeg,.png}
\else

   \usepackage[dvips]{graphicx}
   \graphicspath{{figures/}}
   \DeclareGraphicsExtensions{.eps}
\fi
\usepackage[T1]{fontenc}
\usepackage{amsmath,amssymb,wasysym}
\usepackage{url}
\usepackage{color,soul}
\usepackage{ctable} 
\usepackage{multirow}
\usepackage{array, booktabs, makecell}
\usepackage{lipsum,booktabs}
\usepackage{comment}
\usepackage{threeparttable}
\usepackage{cite}
\usepackage{amssymb}
\ifCLASSOPTIONcompsoc
  \usepackage[caption=false,font=footnotesize,labelfont=sf,textfont=sf]{subfig}
\else
  \usepackage[caption=false,font=footnotesize]{subfig}
\fi

\begin{document}
\bstctlcite{IEEEexample:BSTcontrol} 

\title{LUNA-CIM: Lookup Table based Programmable Neural Processing in Memory}

\author{

    Peyman Dehghanzadeh,
    Baibhab Chatterjee,~\IEEEmembership{Member,~IEEE}
    and Swarup Bhunia,~\IEEEmembership{Senior Member,~IEEE}
    
\thanks{P. Dehghanzadeh, B. Chatterjee and S. Bhunia are with the Department
of Electrical and Computer Engineering, University of Florida, Gainesville,
FL, 32611 USA.

E-mail: \{p.dehghanzadeh, chatterjee.b\}@ufl.edu, swarup@ece.ufl.edu.} }

\maketitle

\begin{abstract}

This paper presents a novel approach for performing computations using Look-Up Tables (LUTs) tailored specifically for Compute-In-Memory applications. The aim is to address the scalability challenges associated with LUT-based computation by reducing storage requirements and energy consumption while capitalizing on the faster and more energy-efficient nature of look-up methods compared to conventional mathematical computations. The proposed method leverages a divide and conquer (D\&C) strategy to enhance the scalability of LUT-based computation. By breaking down high-precision multiplications into lower-precision operations, the technique achieves significantly lower area overheads, up to approximately 3.7 times less than conventional LUT-based approaches, without compromising accuracy. To validate the effectiveness of the proposed method, extensive simulations using TSMC 65 nm technology were conducted. The experimental analysis reveals that the proposed approach accounts for less than 0.1\% of the total energy consumption, with only a 32\% increase in area overhead. These results demonstrate considerable improvements achieved in energy efficiency and area utilization through the novel low-energy, low-area-overhead LUT-based computation in an SRAM array. 

\end{abstract}

\begin{IEEEkeywords}
Compute-In-Memory, Low-energy computation, Look-Up Table, Divide and conquer approach

\end{IEEEkeywords}

\section{Introduction}
\IEEEPARstart{T}{he} exponential growth of data-intensive applications and the demand for faster processing speeds have pushed the boundaries of traditional computing architectures. Artificial Intelligence (AI) and machine learning applications, such as deep learning for image recognition, natural language processing, and recommendation systems, require vast amounts of data for training and inference. The complex calculations involved in training deep neural networks and processing large datasets demand immense computational power and memory bandwidth. As these models become more sophisticated, the need for faster processing speeds and more efficient memory usage becomes increasingly crucial. In response to these challenges, researchers have been exploring innovative approaches to revolutionize the way we perform computations \cite{huang2020superconducting, schuman2022opportunities, 9064516}. One such groundbreaking concept that has garnered significant attention is "Compute in Memory" (CiM)~\cite{yu2021compute} which represents a paradigm shift in computer architecture, aiming to overcome the performance bottlenecks imposed by the traditional von Neumann architecture. Rather than relying solely on the separation of memory and computation, CiM proposes a holistic approach by integrating computation capabilities directly into the memory units~\cite{ielmini2018memory}. By co-locating computing units with memory, CiM offers the potential for substantial gains in terms of computational efficiency, reduced energy consumption, and improved system performance~\cite{yin2020xnor, biswas2018conv}. 

However, there are important challenges that need to be addressed to fully harness its potential. One major challenge is related to scalability. While CiM demonstrates impressive gains on a smaller scale, it's essential to ensure that these benefits can be maintained as systems grow larger and more complex. Connecting a vast number of CiM units while maintaining synchronization and minimizing communication delays is not straightforward. Researchers are investigating scalable interconnect architectures and communication protocols that can uphold the advantages of CiM in larger setups without introducing bottlenecks or compromising efficiency.

This paper presents an innovative approach referred to as LUNA-CiM, aimed at tackling the significant scalability challenge presented by CiM. Our approach delves into practical implementation by offering a tangible strategy for overcoming scalability hurdles. Importantly, our proposed approach also demonstrates a high degree of energy efficiency. This dual focus on both scalability and energy efficiency reflects our commitment to shaping a computing landscape that not only handles increasing demands but also does so in a sustainable and resource-conscious manner. LUNA-CIM at its essence delivers rapid and flexible in-memory processing, coupled with a significant noise margin. Remarkably, these capabilities are harnessed without the requirement for adjustments to existing design practices, simplifying the validation process for FPGA/ASIC applications.

The upcoming sections of this paper are structured as follows to provide a more comprehensive exploration of the suggested approach and its analysis. In Section II, a broad overview of prior efforts in CiM is presented. In Section III, the proposed method for implementing various CiM applications is thoroughly explained. Section IV offers a comprehensive analysis of the proposed architectures. Finally, in Section V, concluding remarks summarize the key findings and contributions of this article.

\section{Prior Research and Developments}

In this section, we provide an overview of the most significant CiM implementations documented in the existing literature. This section provides valuable insights into a variety of approaches along with their respective advantages and disadvantages that have been investigated within the domain of CiM.

\textbf{NVM-based CiM} is an innovative approach that leverages advancements in memory technologies, particularly focusing on emerging non-volatile memory (NVM) devices such as Resistive RAM (RRAM), Phase Change Memory (PCM), and Spin-Transfer Torque MRAM (STT-MRAM)~\cite{xue201924, chen201865nm, mochida20184m, arnaud2018truly, wu201840nm, chih202013, 8976130}. These NVM technologies exhibit unique properties that make them highly suitable for integrating computational capabilities directly within the memory subsystem. These NVM technologies offer significant advantages in terms of high data density, making them capable of storing vast amounts of data in a compact form. Moreover, their non-volatile nature ensures data retention even in the absence of power, rendering them suitable for applications requiring persistent storage. Additionally, NVM devices demonstrate low power consumption, enabling the development of energy-efficient computing systems~\cite{chen2016review}. These NVM technologies hold great potential for highly energy-efficient and low-latency CiM applications. Nonetheless, it's important to recognize that these technologies are still in a relatively early phase of development when compared to traditional memory technologies. Moreover, they encounter a spectrum of reliability hurdles, encompassing issues related to endurance, resilience, and uniformity. Also, they are susceptible to certain types of security vulnerabilities, such as side-channel attacks, due to their physical properties and mechanisms.

\textbf{Analog CiM} is gaining traction as an attractive method. This approach harnesses the inherent characteristics of memory devices to conduct specific computations directly within memory cells, eliminating the need to shuttle data between memory and an independent computing unit (CPU/GPU). This offers potential benefits like accelerated computation for specific assignments and reduced energy usage in comparison to traditional digital methods~\cite{6805187}. Nevertheless, analog CiM has inherent limitations, particularly concerning accuracy in real-world applications~\cite{9365766}. Noise and intrinsic transistor mismatches can compromise precision and reliability in computation within analog CiM. Additionally, the component responsible for converting data between analog and digital domains occupies the largest space and consumes the highest amount of energy within the design. Moreover, Analog CiM suffers from potential write and read latency increases, limited scalability as process nodes shrink, and concerns over endurance, wear, and cell lifetime. These challenges underscore the complex trade-offs involved in implementing analog CiM solutions, necessitating careful consideration and ongoing advancements in analog CiM technology.

\textbf{Digital CiM} is an alternative approach that focuses on performing computations within memory cells, using digital logic circuits. In this method, memory cells are equipped with additional circuitry to enable basic logical and arithmetic operations directly within the memory array, without the need to transfer data to a separate processor. Compared to NVM technologies, Digital CiM based on CMOS technology~\cite{fujiwara20225} is more mature and amenable to large-scale production. Unlike Analog CiM, which utilizes the analog properties of memory cells, Digital CiM relies on well-established digital logic principles. It involves integrating simple computational units, such as adders, multipliers, and shifters, into the memory architecture. This enables the memory cells to perform calculations and process data in a manner akin to conventional digital processors. However, Digital CiM comes with its own set of trade-offs~\cite{8902824}. Digital CiM implemented in CMOS technology requires higher energy consumption compared to NVM-based CiM. Moreover, it incurs a larger area overhead, occupying more physical space on integrated circuits. Furthermore, Digital CiM may suffer from increased compute latency due to factors such as data transfer between memory and processing units and computational pipeline stages. Additionally, Digital CiM faces limitations in computing parallelism due to routing congestion, which can have a substantial impact on overall performance.

\textbf{Lookup-based CiM} is a specialized approach that is used to address the latency challenges in CMOS designs and enable parallelism in Digital CiM~\cite{9251975, 9643667, 10178690, 10090358}. In this approach,  memory cells are employed to store precomputed results or tables, which are then accessed to perform specific computations. Instead of executing arithmetic or logical operations directly within memory cells, lookup-based CiM retrieves predefined values or patterns from memory and uses them to compute results. This method offers several advantages. Firstly, it provides programmable storage, allowing flexibility in configuring the lookup table to meet different computation requirements. Secondly, the lookup-based approach enables fast computation with low latency, as the desired result can be retrieved directly from the lookup table, eliminating the need for complex calculations. Additionally, LUT-based methods exhibit a large noise margin, making them more tolerant to system variations and disturbances, thereby enhancing the reliability and accuracy of the computation. Furthermore, these methods offer the benefits of negligible memory fetch energy and latency since the results are already stored in the lookup table, reducing the need for additional memory fetch operations. Lastly, LUT-based CiM operations have the advantage of simplicity in hardware design, as the lookup table itself serves as the primary computational element, eliminating the need for complex hardware components. 

\begin{figure}[!t]
    \centering
    \includegraphics[width=9cm]{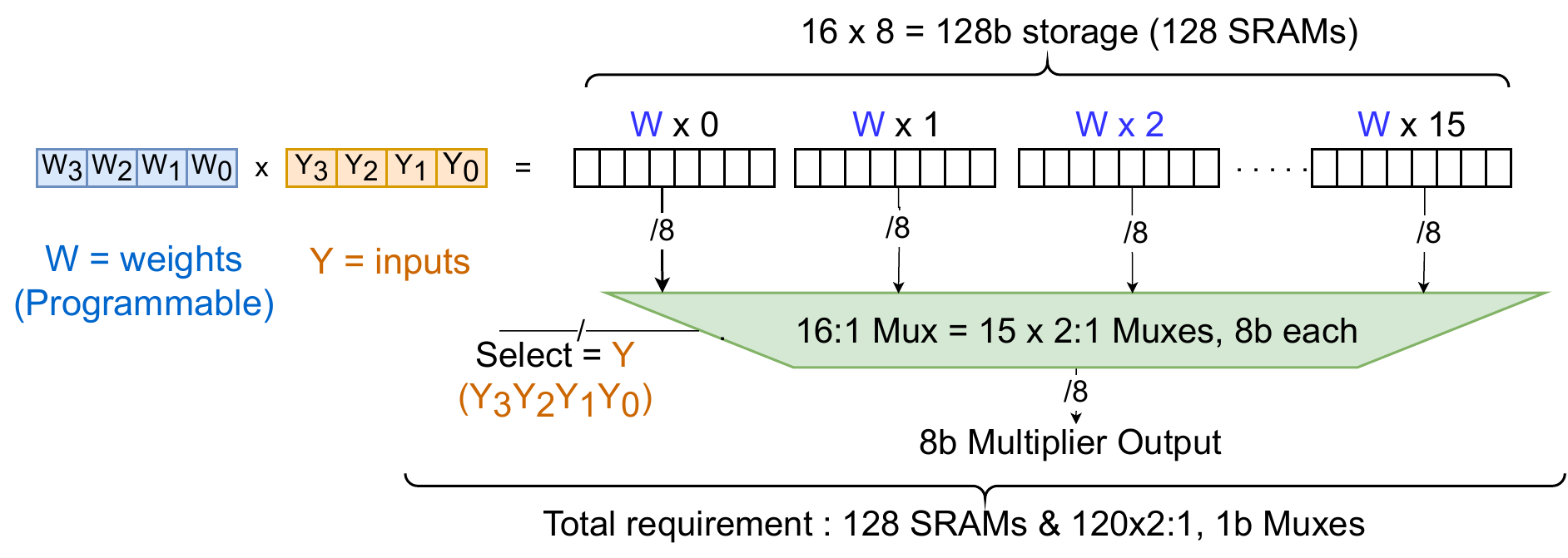}
    \caption{Conventional 4b multiplier using LUT-based technique without optimization.}
    \label{fig:NO_OPTIMIZATION}
\end{figure}

By utilizing lookup table-based methods in Digital CiM, the aforementioned advantages can be utilized to address latency challenges, introduce parallelism, and achieve efficient computation within the memory subsystem~\cite{10178690}. However, it is important to note that traditional LUT-based operations have limitations in terms of scalability. For example, as depicted in Fig.~\ref{fig:NO_OPTIMIZATION}, in a MAC-based neuron, for 4b multiplications (4b weight, W × 4b input, Y), assuming that the weights are already fixed, there are $2^{4}$ different 8b results based on the 4b inputs, which requires $2^{4}$ × 8b = 128b storage. Furthermore, the selection of one result out of the $2^{4}$, is required to have a 16:1, 8b Mux, which is equivalent to 15 instances of a 2:1 Mux for 8b operands, which in turn is equivalent to 120 instances of a 1b, 2:1 Mux. Similarly, for 8b multiplications, we need 4096 SRAM cells, and 4080 instances of the 1b, 2:1 Muxes, making such a LUT-based multiplication unrealistic for CiM. Therefore, the primary obstacle that needs to be addressed before widespread adoption is scalability.

In image and pattern recognition applications, system-level precision requirements often vary based on the specific task and the trade-off between accuracy and computational efficiency. Generally, it has been observed that for many image/pattern recognition tasks, using 4 to 8-bit precision can often be sufficient~\cite{8657364, 8123656, chatterjee2021energy }. Table~\ref{tab:TABLE_1}, shows the required storage and select logic for different multiplication resolutions. A significant escalation as the operand size expands. This growth eventually reaches a point where it becomes impractical and too costly to implement.
\begin{table}[!h]
  \caption{The number of SRAM cells and 2:1 Multiplexers required to perform 3b-8b multiplication using the LUT-based approach.}

  \centering
  \label{tab:TABLE_1}
  \begin{tabular}{ *{3}{c} }
    \toprule
     \thead{Multiplier Bit\\ Resolution} & \thead{Number of SRAMs \\Required}  & \thead{Number of 2:1, 1bit MUXes\\ Required}  \\
    \midrule
          3b    & 48      & 42         \\
          4b    & 128     & 120        \\
          5b    & 320     & 310        \\
          6b    & 768     & 756        \\
          7b    & 1792    & 1778      \\
          8b    & 4096    & 4080       \\
    \bottomrule
  \end{tabular}
\end{table}

The scalability issues faced by LUT-based approaches in handling increasingly larger and more intricate computations necessitate the exploration of alternative strategies and techniques. This is crucial to overcome the limitation and ensure the efficient and scalable implementation of lookup table-based methods in Digital CiM systems. This paper presents a groundbreaking approach for lookup table-based computation, specifically designed for CiM applications. The proposed method addresses the challenges related to scalability in terms of area and storage utilization, as well as energy consumption in LUT-based computation.

\section{Proposed Method}

This section provides a comprehensive description of a divide and conquer (D\&C) based method specifically designed for low-energy, low-area-overhead LUT-based computation in CiM applications. The D\&C approach is leveraged to efficiently address the existing challenges while harnessing the inherent advantages offered by lookup methods. In the D\&C method, the complex problem is decomposed into smaller, more manageable sub-problems, which are then solved independently. The solutions obtained from the sub-problems are subsequently combined to derive the solution for the original problem. By breaking down the problem into smaller components, it becomes feasible to tackle each sub-problem individually, often resulting in improved efficiency and optimized solutions. 

\subsection{LUT-based multiplier using D\&C approach }
\begin{figure}[!t]
    \centering
    \includegraphics[width=9cm]{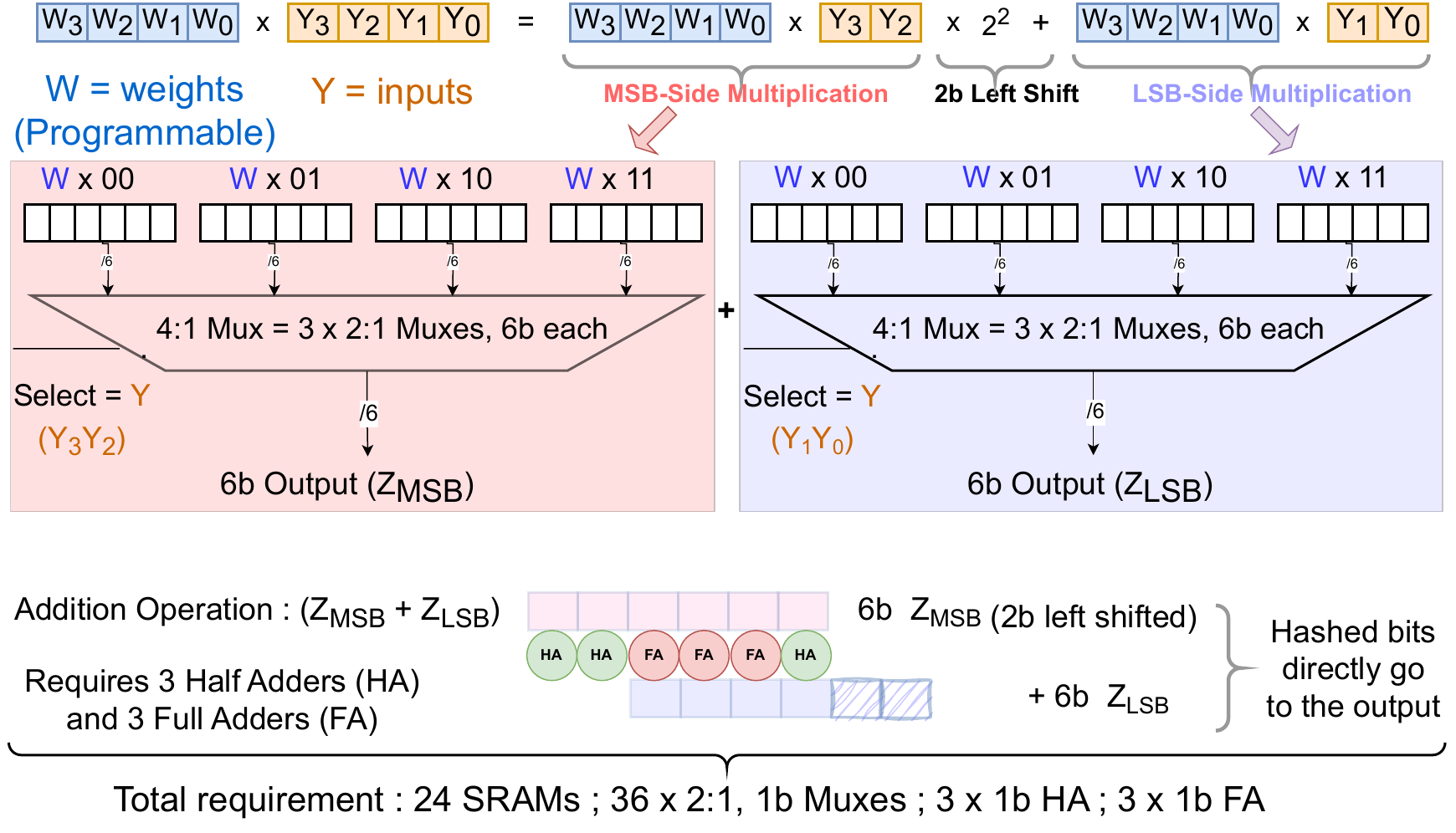}
    \caption{Proposed LUT based 4b multiplier with Divide and Conquer method.}
    \label{fig:DIVIDE_AND_CONQUER}
\end{figure}

The D\&C based approach for LUT-based computation is depicted in the Fig.~\ref{fig:DIVIDE_AND_CONQUER}. In this particular scenario, a 4b × 4b, W × Y multiplication is partitioned into two separate 4b × 2b multiplications. The first multiplication operates on the most significant 2 bits of Y (MSB side), while the second multiplication operates on the least significant 2 bits of Y (LSB side). This division facilitates the efficient distribution of the computation, thereby improving the overall efficiency of the LUT-based computation process. In the next step, the result of the MSB side multiplication ($Z_{MSB}$) undergoes a left shift of 2 bits. This left-shifted result is then added to the result of the LSB side multiplication ($Z_{LSB}$) to obtain the final result of the multiplication. By combining these two partial results, the desired outcome of the multiplication operation is achieved. In the case of each smaller (4b × 2b) multiplication, the conventional approach would typically involve storing a total of 24b (4 possible 6b results, each requiring $2^2$ × 6b of storage). Additionally, a 4:1, 6b Mux would be used, which is equivalent to either 3 instances of a 2:1, 6b Mux or 18 instances of a 2:1, 1b Mux. Also, the addition of the 6b $Z_{MSB}$ (with a 2b left shift) and $Z_{LSB}$ requires 3 instances of 1b half adders (HA) and 3 instances of 1b full adders (FA) due to the left shift operation in $Z_{MSB}$ before the addition. Since 1b FAs are necessary only when there are three inputs, having 3 FAs and 3 HAs will suffice for this purpose. In this specific configuration, a total of 24 storage elements (SRAM cells) are required. It is worth noting that in all the figures presented within this document, these SRAM cells are depicted individually for each Mux. This representation choice is made to ensure clear visualization of the connections and facilitate a better understanding of the overall system. However, it needs to be kept in mind that the number of actual SRAMs will depend on Fanout considerations.

\begin{table*}
  \caption{Comparing Component Complexity: Traditional vs. Optimized Divide \& Conquer LUT-based for 4b-16b Resolution Multiplication.}
  \centering
  \label{tab:TABLE_2}
  \begin{tabular}{c | c c | c c c c}
     \multicolumn{3}{c}{Traditional Approach} & \multicolumn{4}{c}{Optimized D\&C Approach} \\
     \toprule
     \thead{Multiplier Bit\\ Resolution} & \thead{No. of SRAMs \\Required}  & \thead{No. of 2:1, 1bit MUXes\\ Required} & \thead{No. of SRAMs \\Required}  & \thead{No. of 2:1, 1bit MUXes\\ Required} & \thead{No. of HAs \\Required}  & \thead{No. of FAs\\ Required}  \\

    \midrule
          4b    & 128          & 120         & 10    & 36    & 3    & 3\\
          8b    & 4096         & 4080        & 36    & 120   & 11   & 21 \\
          16b   & 2097152      & 2097120     & 136   & 432   & 31   & 105  \\

    \bottomrule
  \end{tabular}
\end{table*}

\subsection{Optimized LUT-based multiplier using D\&C approach }
\begin{figure}[!b]
    \centering
    \includegraphics[width=9cm]{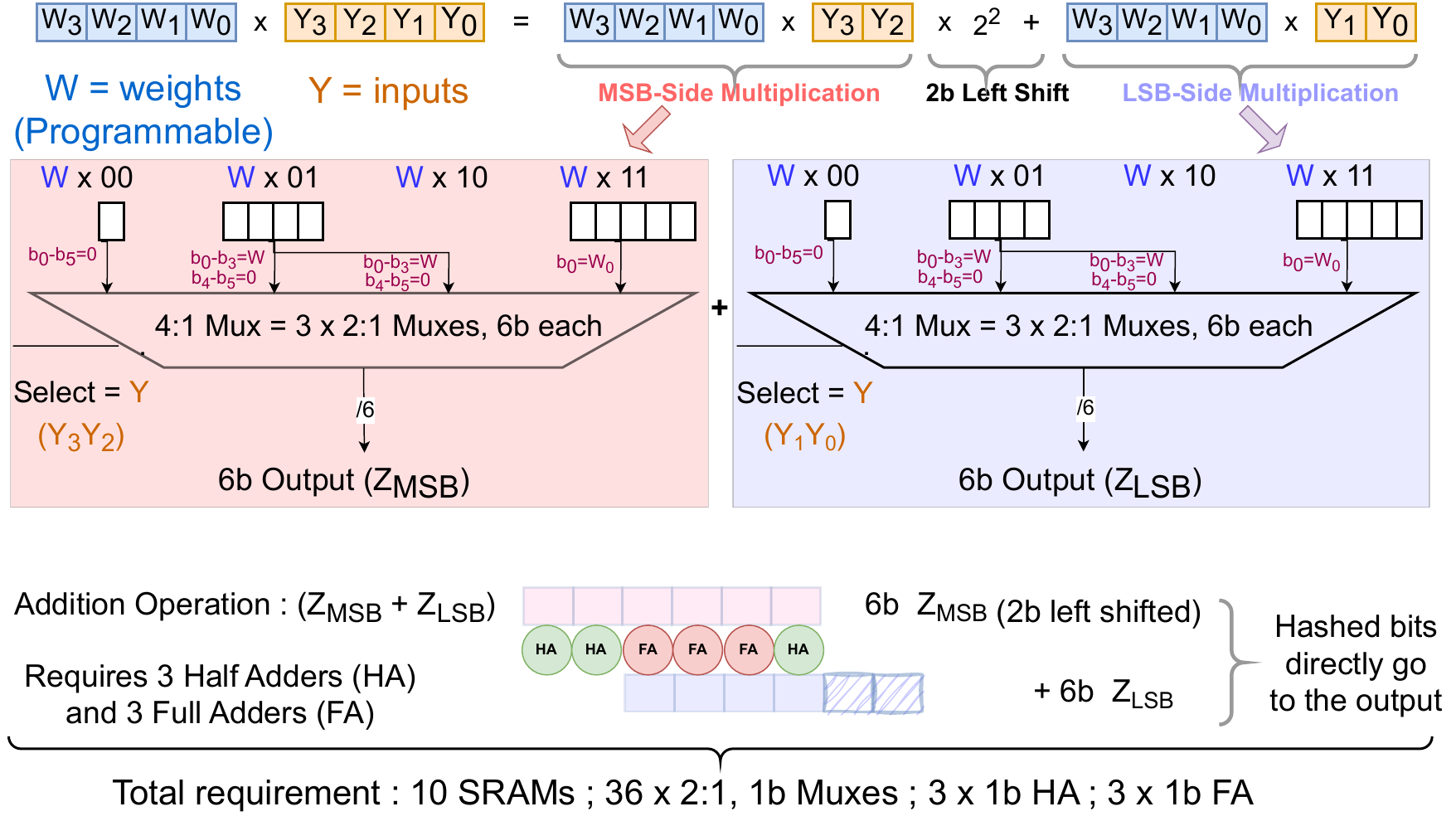}
    \caption{Optimized LUT based 4b multiplier with Divide and Conquer method.}
    \label{fig:OPT_DIVIDE_AND_CONQUER}
\end{figure}

In the proposed idea, for a 4b × 4b, W × Y multiplication the number of storage elements has significantly decreased from 128 to 24. This reduction in the required storage elements showcases a substantial improvement achieved in the design. However, there is an opportunity for optimizing the required number of SRAM cells for storage. In the case of the smaller (4b × 2b) multiplications, as depicted in Fig.~\ref{fig:OPT_DIVIDE_AND_CONQUER}, the weight W (4b) can only be multiplied with any of the combinations 00, 01, 10, and 11 for the 2b operand. In the case of the multiplication W × 00, only one bit (0) needs to be stored, which will be connected to all six bits of one of the inputs of the 4:1 Mux. This optimized approach significantly reduces the storage requirements for this specific multiplication. When performing the multiplication W × 01, it is sufficient to store only the four bits of W. These four bits will be connected to the four LSBs of another distinct 6b input of the 4:1 Mux. Meanwhile, the two MSBs will be connected to 0, resulting in an optimized storage configuration for this particular multiplication. In the case of the multiplication W × 10, there is no need to store any additional bits. Instead, the stored 4b result of W × 01 can be left-shifted by one bit and connected to the four middle bits of another distinct 6b input of the 4:1 Mux. The MSB and LSB of this input will be connected to 0. Finally, for the multiplication W × 11, it is necessary to store the five MSBs of the result. These five MSBs will be connected to the corresponding five MSBs of the final distinct 6b input of the 4:1 Mux. Additionally, the least significant LSB of the 6b input will be connected to the LSB of W, which was already stored for the operation W × 01. This storage configuration optimizes the representation for the W × 11 multiplication. By employing this optimization approach, the total configuration requires 10 SRAM cells, 36 instances of a 2:1, 1b Mux, 3 FAs, and 3 HAs. This optimized setup efficiently accommodates the required components for the desired operations.

Additionally, it is important to highlight the applicability of this optimization across a spectrum of bit resolutions. Table~\ref{tab:TABLE_2} showcases the tangible results stemming from the implementation of this optimization technique, outlining the essential components required for each specific resolution. Through a side-by-side comparison of the traditional approach and the optimized alternative, this tabular representation vividly illustrates the efficacy of the optimized D\&C methodology in directly tackling the scalability obstacles often encountered within LUT-based strategies.

\subsection{LUT-based multiplier using approximating D\&C approach}

\begin{figure}[!b]
    \centering
    \includegraphics[width=9cm]{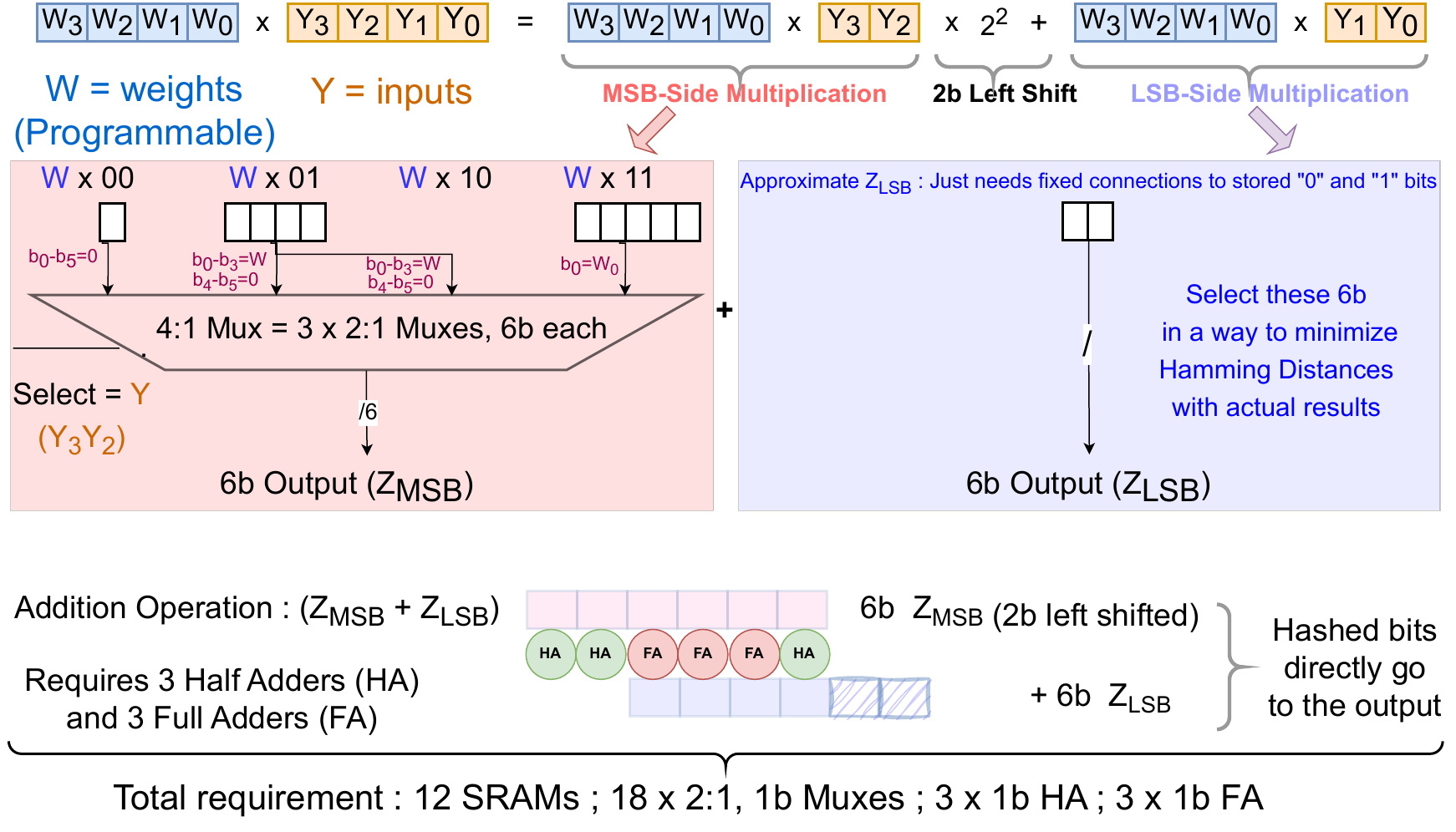}
    \caption{Simplified D\&C approach by approximating the LSB-side multiplication’s result to a fixed value for a LUT based 4b multiplier. }
    \label{fig:APPROX}
\end{figure}

The D\&C structure presented in Fig.~\ref{fig:OPT_DIVIDE_AND_CONQUER} can be further simplified by approximating the result of the LSB-side multiplication with a fixed value. This alternative approach is referred to as ApproxD\&C. This technique is particularly valuable in certain approximate computing problems, such as those encountered in neuromorphic computing with a high error tolerance. In these scenarios, as depicted in Fig.~\ref{fig:APPROX}, it is possible to disregard or approximate multiple LSBs without experiencing a significant decline in the accuracy of the operation's result or the final result of the overall application. This allows for improved efficiency and faster processing while maintaining an acceptable level of accuracy for the specific problem domain. In this particular situation, the selection of a fixed $Z_{LSB}$ involves choosing a specific value that minimizes the Hamming distance between the selected $Z_{LSB}$ and the potential values of $Z_{LSB}$. This selection process aims to optimize the accuracy of the approximation while minimizing any loss in precision.

Since the MSB side operation requires the same hardware, while the LSB side operation only necessitates a specific combination of 0s and 1s, the overall hardware requirements can be optimized. Specifically, the LSB side can be simplified by utilizing only 2 bits of storage and eliminating the need for a Mux altogether. As a result, the total hardware requirement for the system is reduced to 12 SRAMs, 18 instances of a 2:1, 1b Mux, 3 instances of 1b HA, and 3 instances of 1b FA. This optimized configuration ensures efficient utilization of resources while maintaining the desired functionality of the system.

\begin{figure}[!t]
    \centering
    \includegraphics[width=1\columnwidth]{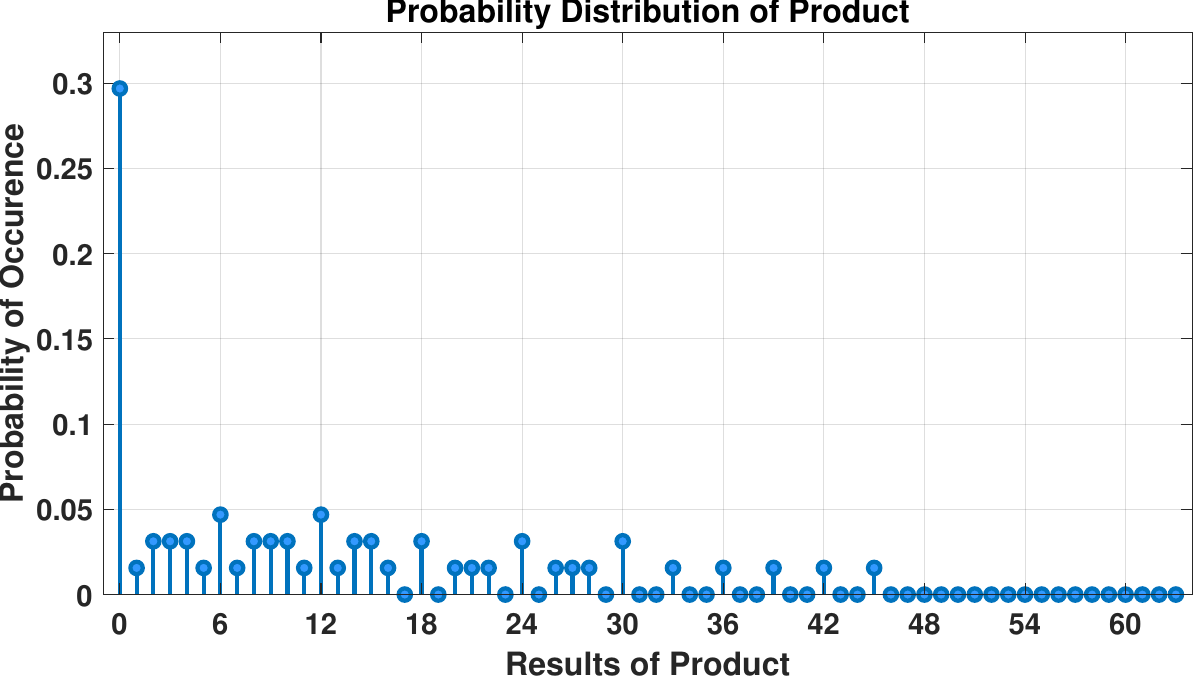}
    \caption{Probability Distribution of (4b × 2b) LSB-Side Multiplication. The plot illustrates the probabilities of individual numbers within the range of 0-63 resulting from the multiplication of two binary numbers. Operand 1 is randomly chosen from [0-15], and operand 2 is randomly chosen from [0-3], both with equal probability. The distribution reveals a notable likelihood of the product being 0, driven by the range of operand values.}
    \label{fig:PROB}
\end{figure}

The analysis reveals that certain values within the range of 0-63 cannot be the result of the (4b × 2b) LSB-side multiplication. These values include 17, 19, 23, 25, 29, 31, 32, 34, 35, 37, 38, 40, 41, 43, 44, and 46-63. This information is illustrated in the stem chart displayed in Fig.~\ref{fig:PROB}, which depicts the probabilities of individual numbers within the range of 0-63 being the outcome of the LSB-side multiplication.

\begin{figure}[!h]
    \centering
    \includegraphics[width=1\columnwidth]{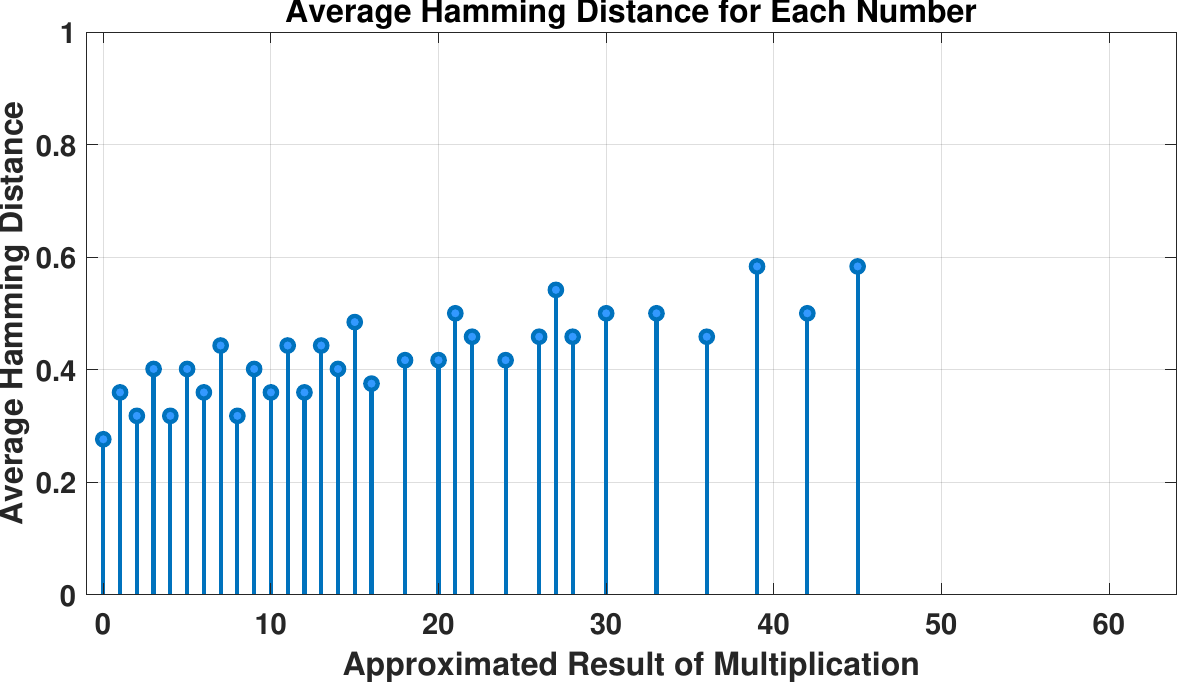}
    \caption{Hamming Distance Analysis: Evaluating the Accuracy of (4b × 2b) LSB-Side Multiplications. This graph depicts the average Hamming distances for each possible number when compared to the approximated results of the (4b × 2b) multiplications. Notably, the lowest Hamming distance of 0.275 is obtained when the approximated value of the multiplication is 0}
    \label{fig:HAM}
\end{figure}

A thorough investigation was undertaken, employing the Hamming distance as a measure, to assess the precision of the estimated results derived from the (4b × 2b) LSB-side multiplications. Hamming distance provides a measure of dissimilarity or discrepancy between two binary strings by counting the differing bits. In Fig.~\ref{fig:HAM}, which shows the results of this analysis, the x-axis represents a range of possible 6-bit approximated results, spanning from 0 to 63 (represented as binary values from 000000 to 111111). On the other hand, the y-axis represents the average Hamming distance. The analysis reveals that an approximated value of $Z_{LSB}=0$ exhibits the lowest Hamming distance. This can be attributed to the fact that the probability of the (4b × 2b) LSB-side multiplication resulting in 0 is the highest among the possible 63 values.

\begin{figure}[t]
    \centering
    \includegraphics[width=1\columnwidth]{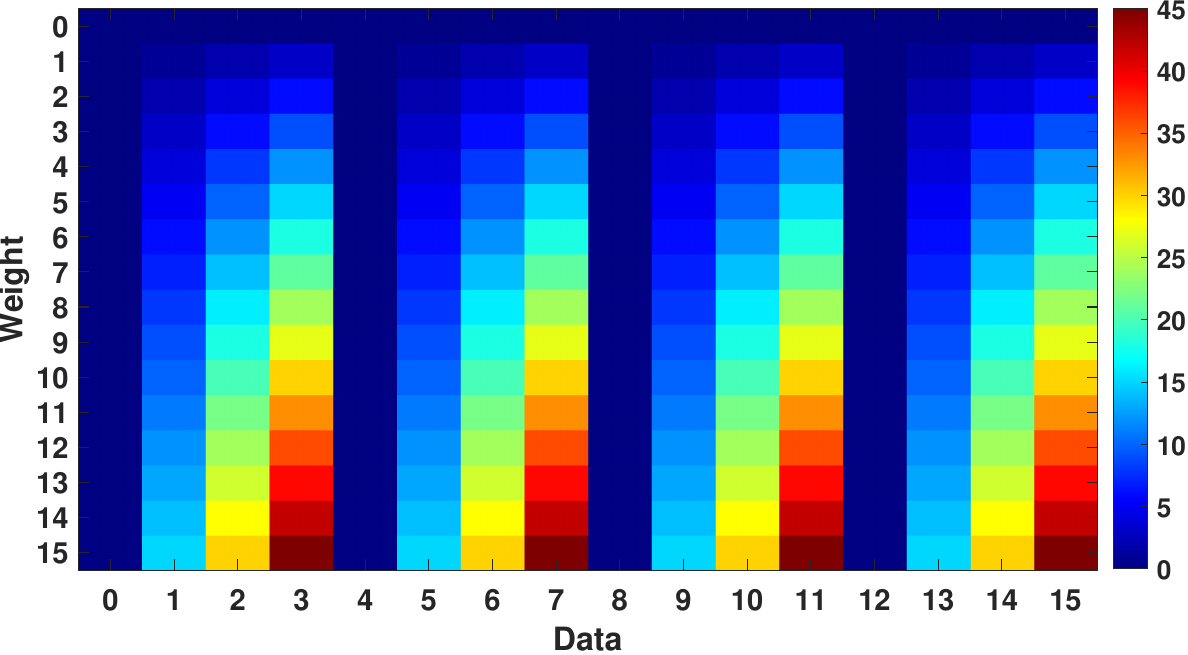}
    \caption{ Heatmap illustrating the relationship between Data (x-axis) and Weight (y-axis), with color intensity indicating the disparity between the D\&C approach and ApproxD\&C results.}
    \label{fig:HEATMAP_1}
\end{figure}

 Since 0 has the highest probability of occurrence (0.296), the average hamming distance for $Z_{LSB}$ (approx) = 0 will be minimum. Fig.~\ref{fig:HEATMAP_1} displays a heatmap in which the x-axis corresponds to Data, the y-axis corresponds to Weight, and the color intensity reflects the variance between the D\&C approach and ApproxD\&C outcomes. The color spectrum ranges from warmer hues indicating greater disparities to cooler shades denoting smaller divergences. This visualization provides a straightforward means of comprehending the variations across all conceivable weight and data combinations.

 \begin{figure}[!t]
    \centering
    \includegraphics[width=1\columnwidth]{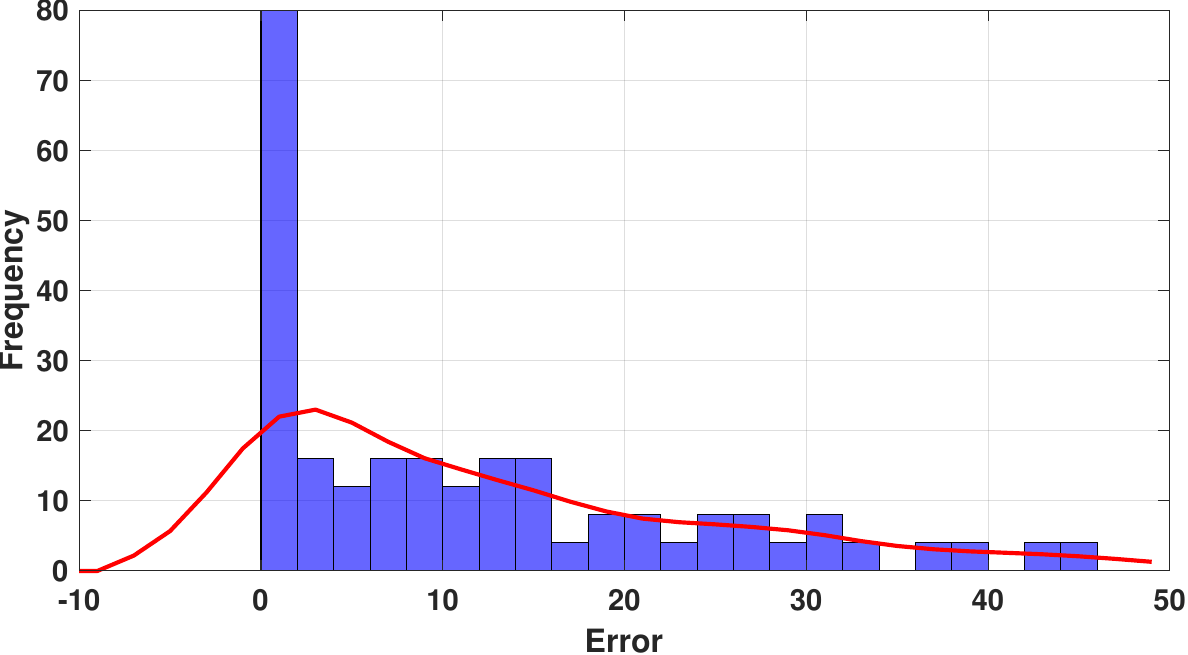}
    \caption{Histogram showcasing Error differences between the D\&C approach and ApproxD\&C results shown in Fig.~\ref{fig:HEATMAP_1}. The x-axis reveals the error range, while the y-axis quantifies occurrence frequency. }
    \label{fig:HISTOGRAM_1}
\end{figure}

In Fig.~\ref{fig:HISTOGRAM_1}, a histogram provides a comprehensive representation of the distribution of Error (Differences between the D\&C approach and ApproxD\&C results) illustrated in Fig.~\ref{fig:HEATMAP_1}. The x-axis showcases the range of Errors (0 to 45) and the y-axis highlights the frequency of occurrences, offering insights into the prevalence of distinct differences. By analyzing the distribution pattern, it becomes possible to pinpoint typical ranges of errors and to grasp the prevalence of specific error magnitudes, contributing to a more nuanced interpretation of the data.

\begin{figure}[!t]
    \centering
    \includegraphics[width=9cm]{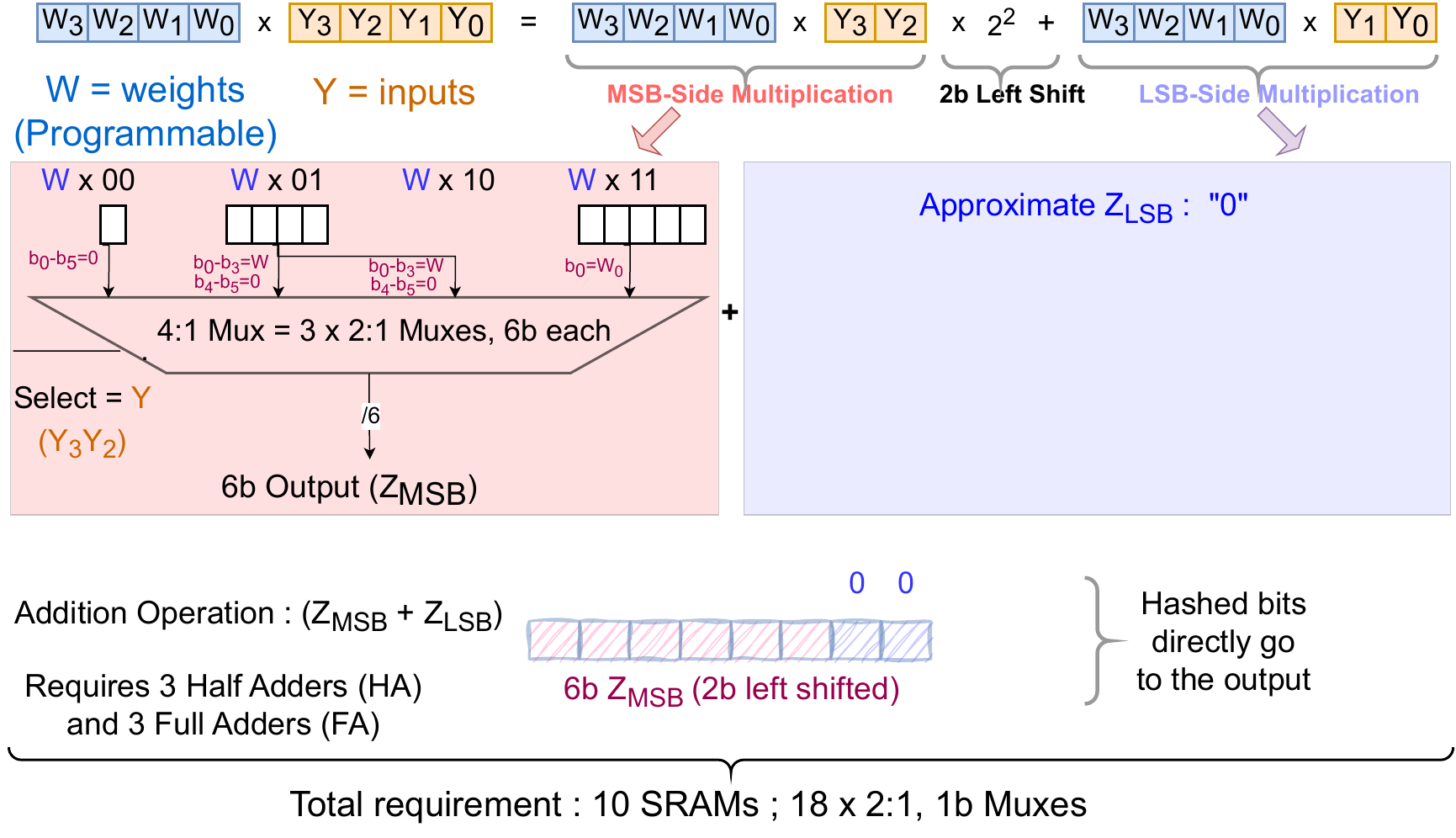}
    \caption{Final structure for the ApproxD\&C approach. In this structure, the $Z_{LSB}$ is set to 0 to reduce overall area overhead.}
    \label{fig:FINAL_APPROX}
\end{figure}

 The finalized structure for the ApproxD\&C approach is depicted in Fig.~\ref{fig:FINAL_APPROX}. As the approximation sets $Z_{LSB}$ to 0, no additional storage or Muxing is necessary for $Z_{LSB}$. Moreover, since $Z_{LSB}$ is approximated to be 0, there is no need to add anything to $Z_{MSB}$ either. Consequently, the overall hardware requirement for this version of ApproxD\&C is reduced to 10 SRAMs and 18 instances of a 2:1, 1b Mux. There is no requirement for HA and FA in this simplified version of the approach.

\begin{figure}[!t]
    \centering
    \includegraphics[width=1\columnwidth]{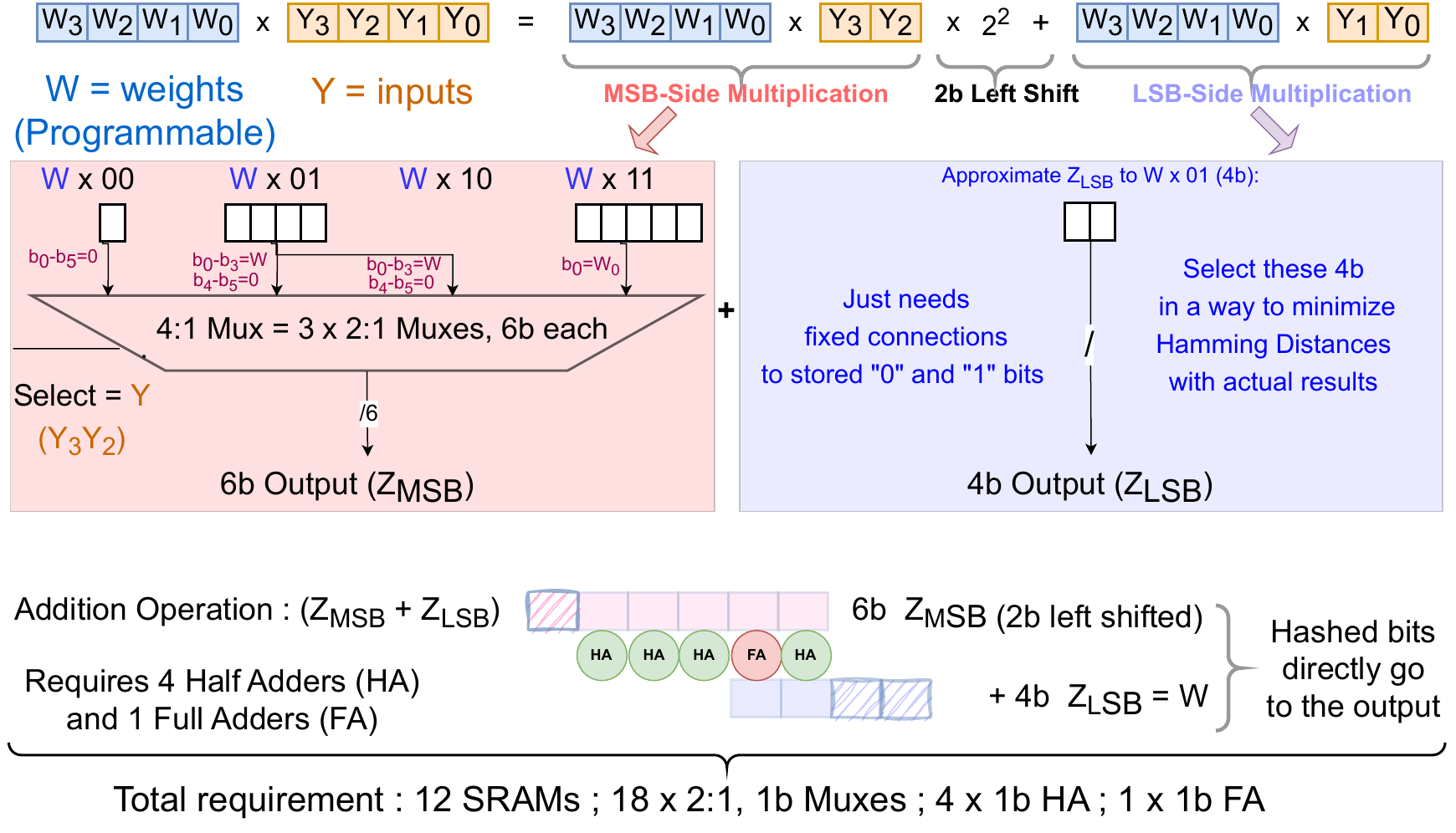}
    \caption{ApproxD\&C 2 an alternative implementation for the ApproxD\&C approach.}
    \label{fig:ALT}
\end{figure}

\begin{figure}[!t]
    \centering
    \includegraphics[width=1\columnwidth]{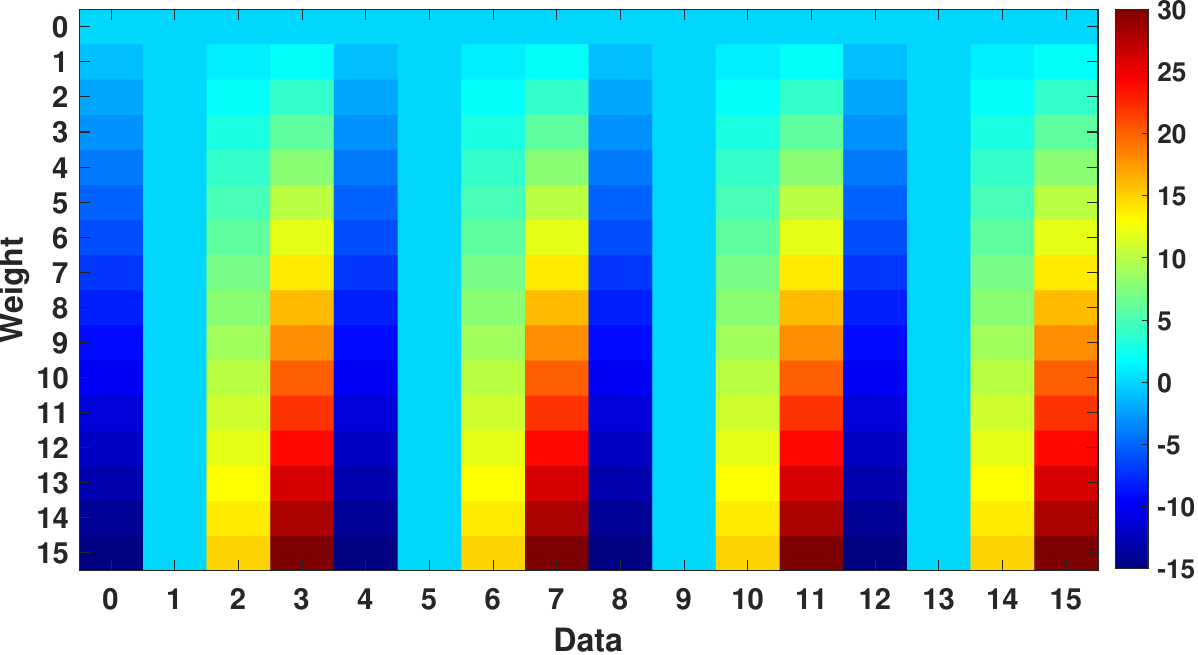}
    \caption{Heatmap illustrating the relationship between Data (x-axis) and Weight (y-axis), with color intensity indicating the disparity between the D\&C approach and ApproxD\&C 2 results.}
    \label{fig:HEATMAP_2}
\end{figure}

\begin{figure}[!t]
    \centering
    \includegraphics[width=1\columnwidth]{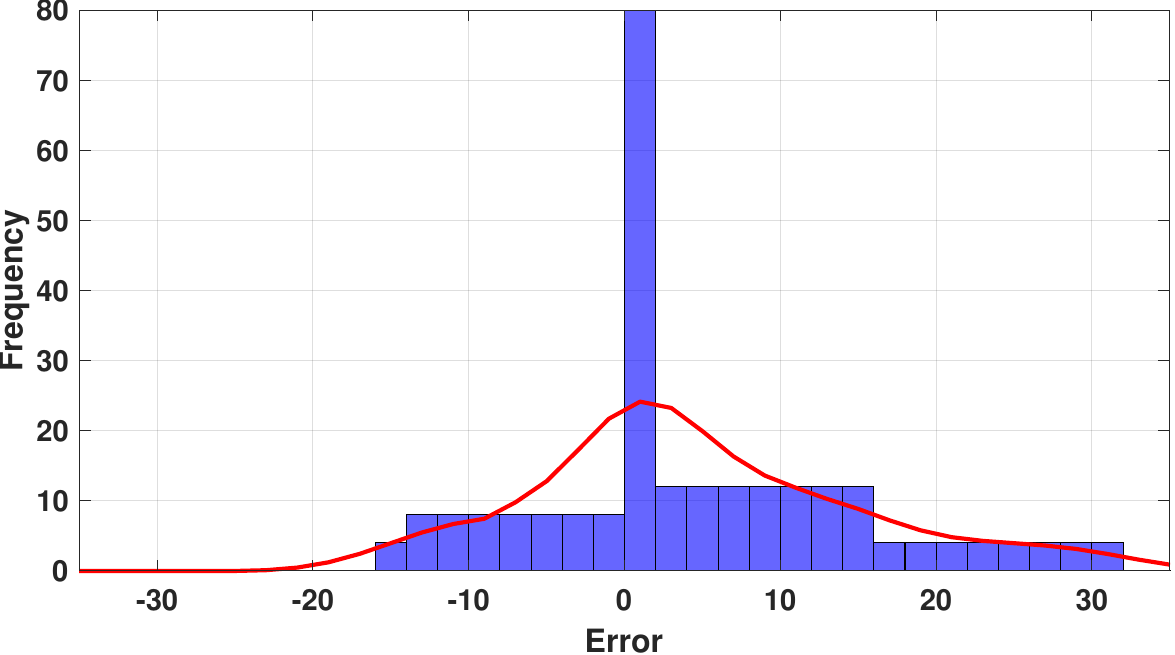}
    \caption{Histogram showcasing Error differences between the D\&C approach and ApproxD\&C 2 results shown in Fig.~\ref{fig:HEATMAP_2}. The x-axis reveals the error range, while the y-axis quantifies occurrence frequency.}
    \label{fig:HISTOGRAM_2}
\end{figure}

There are alternative implementations of the ApproxD\&C approach. As illustrated in Figure~\ref{fig:ALT}, a variant known as ApproxD\&C 2 entails approximating $Z_{LSB}$ as a function of W. In this scenario, the 4-bit W value itself can serve as an approximation for $Z_{LSB}$. Specifically, the 4 LSBs of $Z_{LSB}$ can be directly connected to W, while the 2 MSBs can be connected to 0. Fig.\ref{fig:HEATMAP_2} exhibits a heatmap with Data on the x-axis, Weight on the y-axis, and color intensity representing the difference between the D\&C approach and ApproxD\&C 2 outcomes. In Fig.\ref{fig:HISTOGRAM_2}, the comprehensive histogram mirrors the distribution of Error (Differences) illustrated in Fig.~\ref{fig:HEATMAP_2}. The x-axis spans Errors (-15 to 30), and the y-axis shows occurrence frequency, revealing insights into distinct differences' prevalence. In essence, the balanced error distribution in ApproxDC 2 enhances its versatility across diverse computational contexts. This balance empowers designers with increased flexibility, precision, and adaptability, making ApproxDC 2 a potent solution for navigating the intricate interplay between accuracy and efficiency.

As the maximum value of $Z_{MSB}$ is "101101," it results in the carry-out of the most left HA always being 0. Consequently, there is no carry propagation necessary in the HA. Hence, the output of the MSB of $Z_{MSB}$ can be directly obtained without any additional computations or logic. This simplification of the design minimizes hardware complexity and improves the efficiency of the implementation. This structure requires 12 SRAMs, 18 instances of a 2:1, 1b Mux, 4 instances of 1b HA and 1 instance of 1b FA.

\section{Analysis of the Proposed Technology} 
In this section, we aim to present a comprehensive comparison among the traditional LUT-based CiM implementation, and different D\&C LUT-based CiM implementations. Our comparison focuses on three key factors: area overhead, energy consumption, and accuracy. These factors are crucial in evaluating the efficiency and effectiveness of the different CiM implementations.


\subsection{Accuracy}
\begin{figure}[t]
\centering
\begin{tabular}{@{}c@{}}
\includegraphics[width=.51\linewidth]{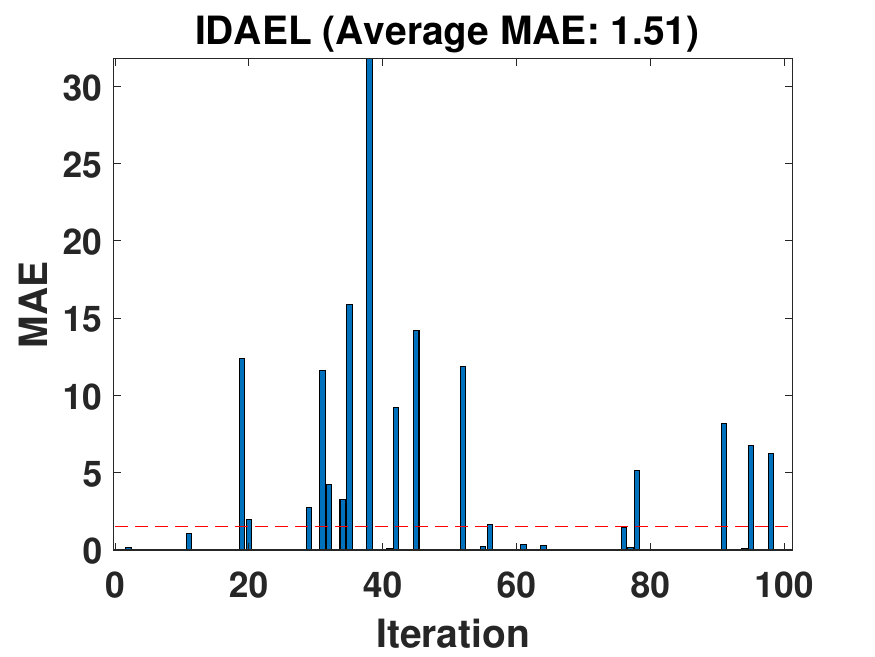}
\includegraphics[width=.51\linewidth]{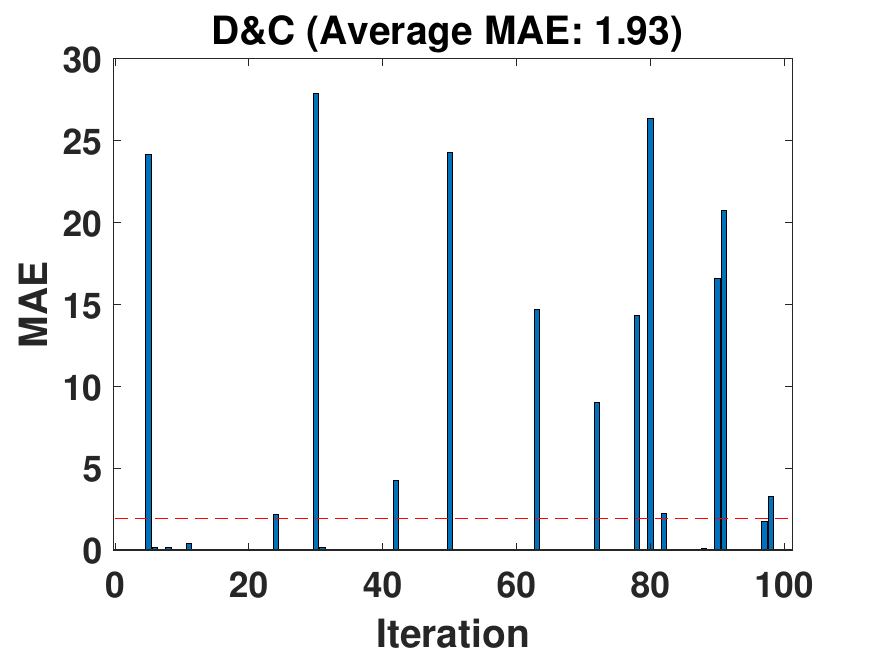}\\[8pt]
\includegraphics[width=.51\linewidth]{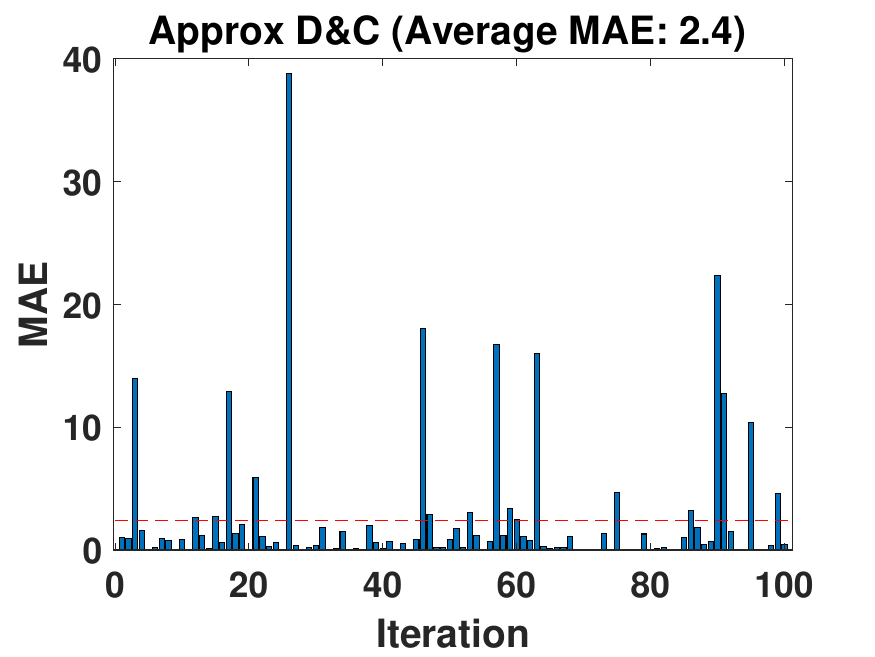}
\includegraphics[width=.51\linewidth]{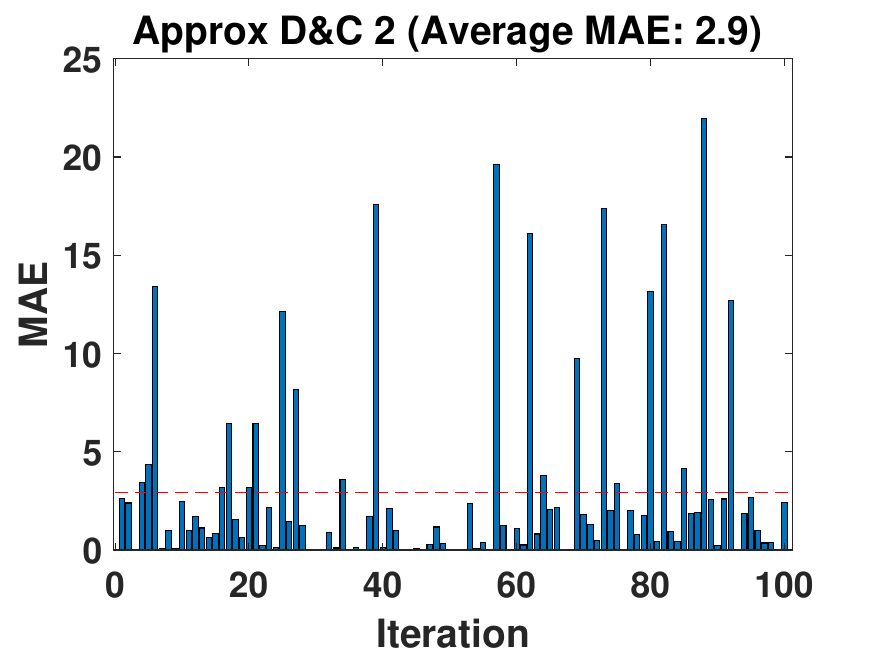}
\end{tabular}%
\caption{\small Mean Absolute Error (MAE) for different D\&C configurations. }
\label{fig:MAE}
\end{figure}

A MATLAB script was developed to explore the functioning of specialized multiplication methods such as D\&C, Approx D\&C, and Approx D\&C 2 within neural networks. We compared their performance to that of "IDEAL" multiplication for accuracy. Random input data was used to evaluate their accuracy by calculating the Mean Absolute Error (MAE) across 100 iterations. These functions are designed to operate on pairs of 4-bit numbers, producing 8-bit outcomes. They were integrated into neural networks for approximation purposes. Our primary objective was to assess how well these specialized multiplication methods align with actual results, measured through the MAE metric. By meticulously comparing the neural networks' outputs with the true outcomes generated by these methods over numerous rounds, we gained insight into their efficacy. As shown in Fig. \ref{fig:MAE}, these results, displayed via Mean Absolute Error (MAE) values and bar charts, provide insights into the efficiency of each approach within this innovative framework. We designed separate neural networks for each method, and subjected them to training and testing, aiming to comprehend their level of accuracy.

\subsection{Energy Consumption}
\begin{figure}[!t]
    \centering
    \includegraphics[width=1\columnwidth]{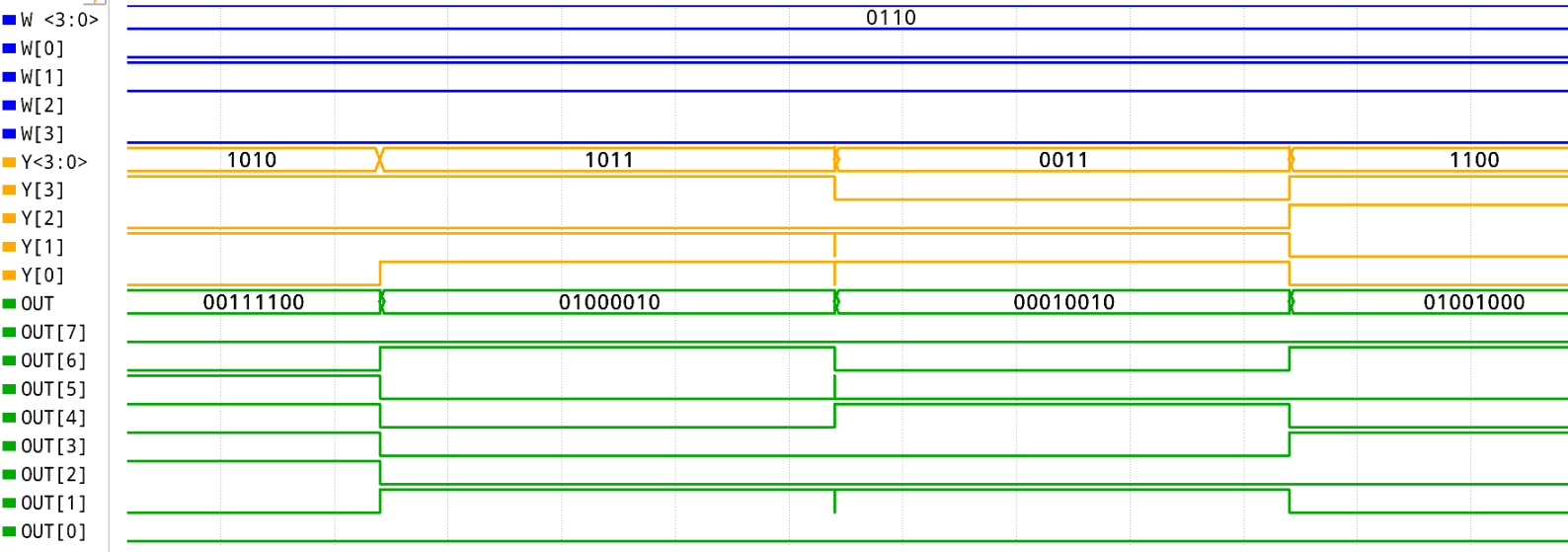}
    \caption{The figure demonstrates the transient simulation of the multiplier output, where the value of W <3:0> remains constant at 0110, while four different Y <3:0> values (1010, 1011, 0011, and 1100) are sequentially applied to the multiplier.}
    \label{fig:TRAN_A}
\end{figure}

The LUNA-CiM concept  was put into practice by integrating it into an 8x8 SRAM array built using TSMC 65 nm technology. This specific SRAM array is composed of 64 SRAM cells and includes 8 units for Bitline conditioning, 8 sense amplifiers, 8 column controllers, as well as a row decoder, a column decoder, and a 4b resolution Mux-based multiplier. The SRAM array was configured to store a fixed value of W <3:0> = 0110. Additionally, we sequentially stored four distinct Y <3:0> values of 1010, 1011, 0011, and 1100 within the array. To evaluate the performance of our design, we incorporated a 4:1 multiplexer. During each discrete time period, we selected one of the aforementioned Y values and routed it to the LSB-side of the mux-based multiplier. This setup enabled us to assess the efficiency and effectiveness of our structure in executing computations based on the specific Y value chosen. As a result of applying different Y values to the LSB side of the multiplier, the output OUT <7:0> exhibited corresponding changes. Fig.~\ref{fig:TRAN_A} visually represents the final results obtained by utilizing the fixed W value and varying Y values.

Energy measurement plays a crucial role in evaluating the energy efficiency of SRAM arrays during data write operations. In this context, energy consumption is commonly quantified in joules per bit per access, providing a precise measure of the energy expended when writing a single bit of data within the memory array. By assessing energy usage at the bit level during write operations, this metric facilitates a detailed analysis of the energy efficiency of SRAM-based systems. The measurement results indicate that the energy per bit per access for the 8x8 SRAM array is 173.8E-12. In comparison, the MUX-based multiplier has an energy consumption share of 47.96E-15, which accounts for approximately 0.0276\% of the overall energy consumption of the SRAM array. This finding highlights the relatively small contribution of the MUX-based multiplier to the total energy usage of the system.

\begin{figure}[!h]
    \centering
    \includegraphics[width=1\columnwidth]{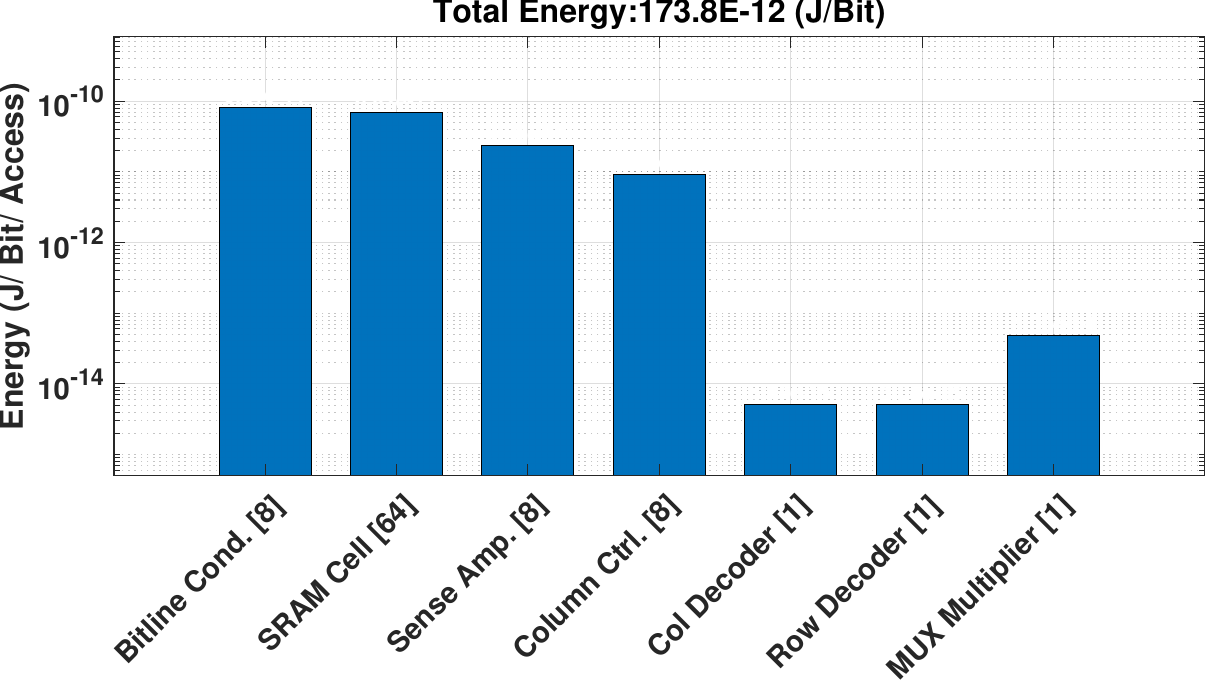}
    \caption{Energy consumption analysis of main components in an 8x8 SRAM Array at TSMC 65 nm technology and 27°C temperature}
    \label{fig:ENERGY_B}
\end{figure}

The bar chart presented in Fig.~\ref{fig:ENERGY_B} compares the energy consumption of the main components within the 8x8 SRAM array. This visualization demonstrates the energy efficiency of the proposed LUNA-CiM approach, showcasing its advantages in addressing various challenges, including scalability, memory fetch energy, and latency. By analyzing the energy consumption of each component, the bar chart highlights how the LUNA-CiM approach offers significant improvements in energy efficiency compared to traditional methods. This comprehensive evaluation underscores the potential of the LUNA-CiM approach not only in solving existing challenges but also in effectively managing energy consumption within SRAM arrays.

\subsection{Area Overhead}
\begin{figure}[!t]
    \centering
    \includegraphics[width=1\columnwidth]{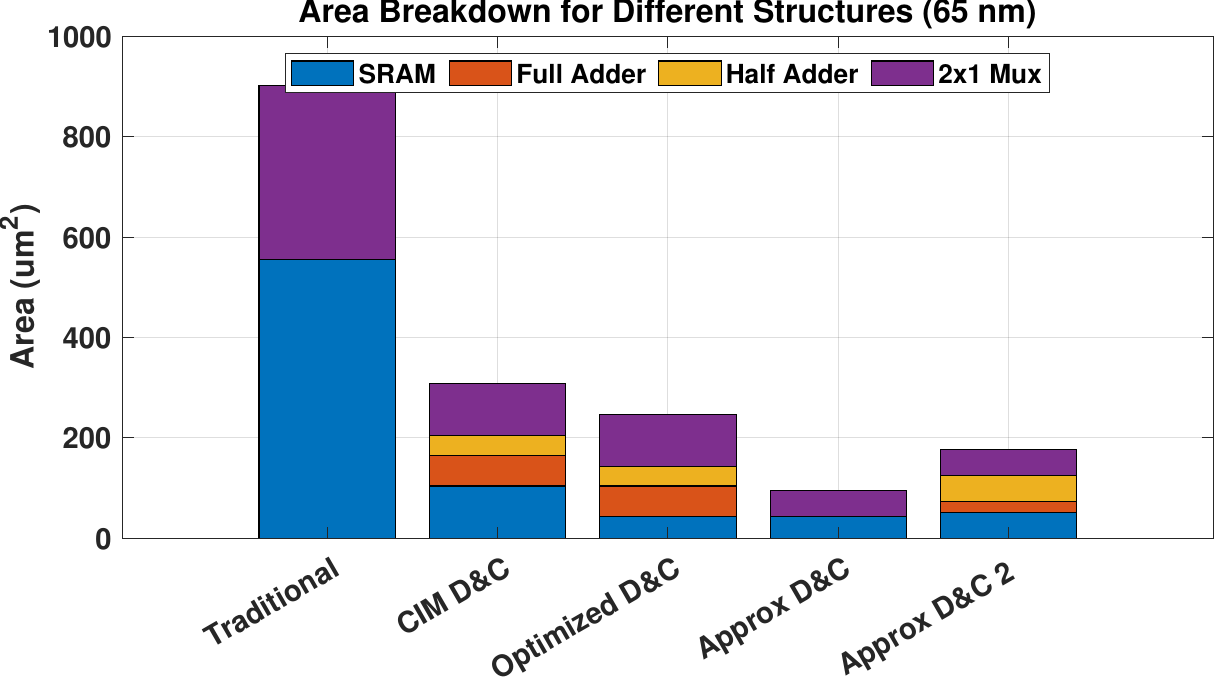}
    \caption{Area overhead comparison among different configurations for 4b
weight, W × 4b input, Y.}
    \label{fig:AREA_A}
\end{figure}

Fig.~\ref{fig:AREA_A} illustrates the measured die area consumption of different solutions, including traditional LUT-based multiplier, D\&C, Optomized D\&C, ApproxD\&C, and ApproxD\&C 2 (Figures:~\ref{fig:NO_OPTIMIZATION}, ~\ref{fig:DIVIDE_AND_CONQUER}, ~\ref{fig:OPT_DIVIDE_AND_CONQUER},~\ref{fig:APPROX} and~\ref{fig:ALT} respectively). For the purpose of a meaningful comparison, the TSMC 65 nm digital library was employed as a reference to determine the count of PMOS/NMOS transistors in each configuration. Compared to the traditional LUT-based CIM implementation, the D\&C approaches demonstrate a significant area benefit, while the ApproxD\&C approach showcases an even greater area benefit. The graph along with the individual segments within each bar it contains offers valuable information regarding the proportional area occupied by each of the components. Additionally, the graph illustrates that even when employing standard cells for FAs and HAs, their respective area utilization is not considerable. These results highlight the potential for significant area optimizations by adopting the D\&C, and ApproxD\&C approaches in contrast to the traditional LUT-based CIM implementation.

The proposed LUNA-CiM approach introduces an area overhead to the SRAM array, which is directly proportional to the number of LUNA-CiM units added. As more LUNA-CiM units are incorporated into the SRAM array, the area overhead increases accordingly. This area overhead accounts for the additional space required to accommodate the LUNA-CiM units and their associated components. To assess the maximum area overhead introduced by the LUNA-CiM approach, four separate LUNA-CiM units were added to the 8x8 SRAM array. 

 \begin{figure}[!h]
    \centering
    \includegraphics[width=0.9\columnwidth]{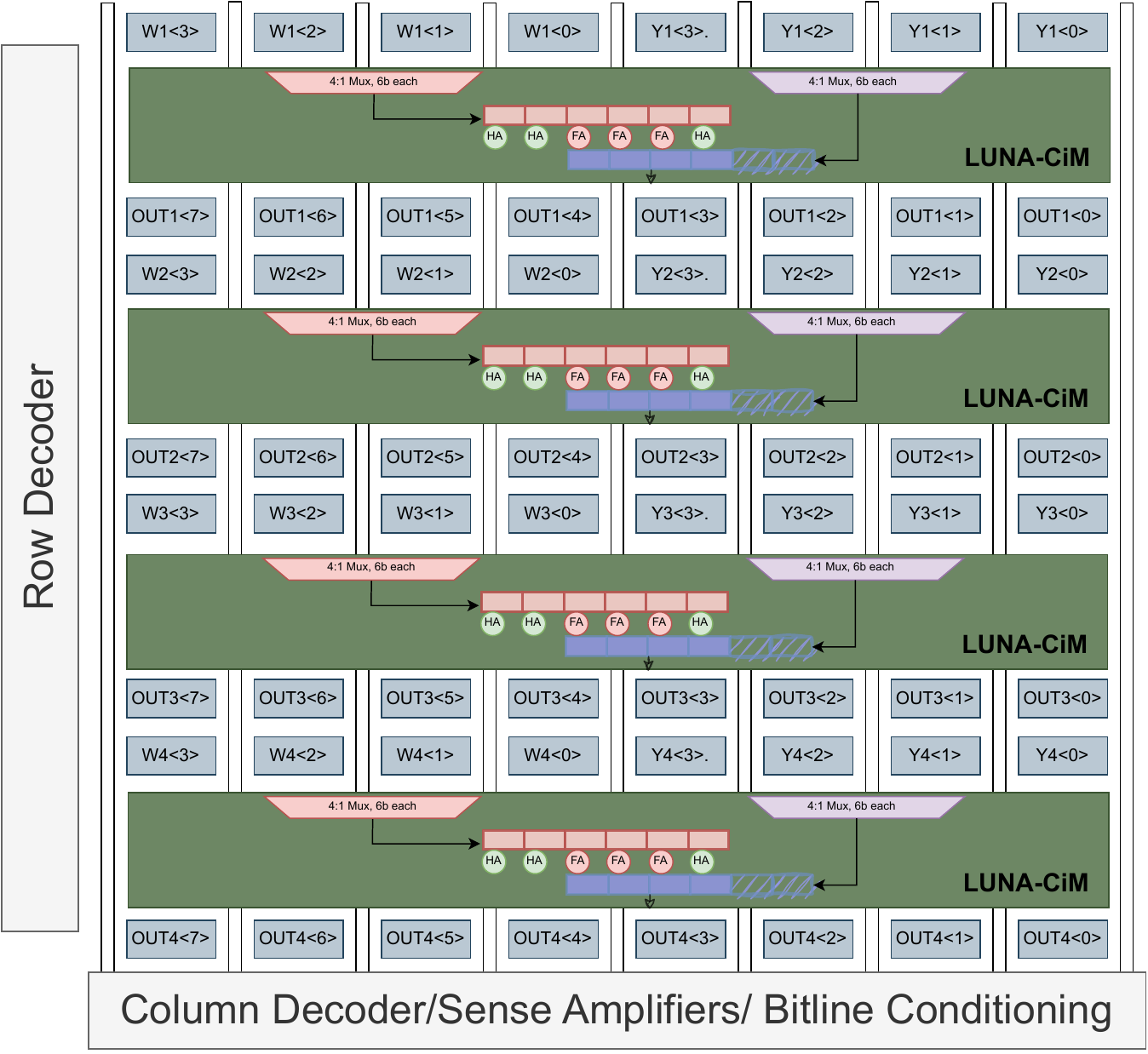}
    \caption{ 8x8 SRAM array featuring the integration of four LUNA-CiM units. Each unit processes inputs from the upper row, performs multiplication, and transmits results to the lower row.}
    \label{fig:SRAM}
\end{figure}

\begin{figure}[!t]
    \centering
    \includegraphics[width=0.8\columnwidth]{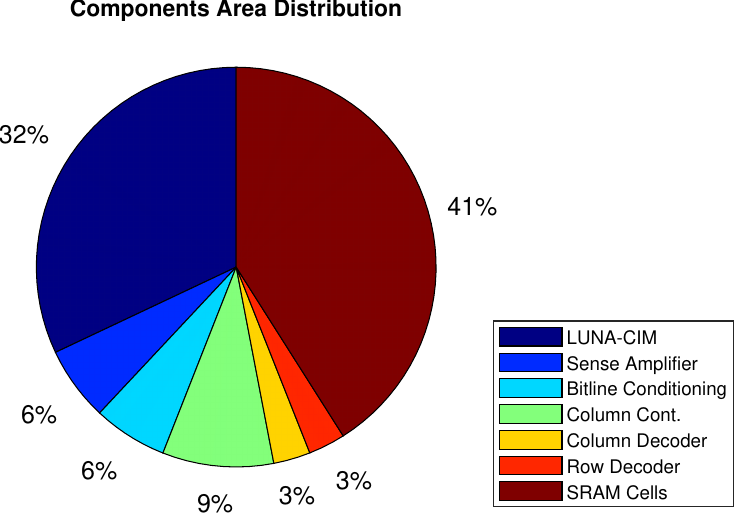}
    \caption{Pie Chart Illustrating Area Overhead in an 8x8 SRAM Array with Four Sets of LUNA-CiM}
    \label{fig:PIE}
\end{figure}

Fig.~\ref{fig:SRAM} illustrates the specific configuration employed: the first LUNA-CiM unit was inserted between the first and second rows, taking inputs (W and Y) from the upper row and delivering the computed results to the lower row. This implementation pattern was replicated for the second, third, and fourth LUNA-CiM units. Fig.~\ref{fig:PIE} displays a pie chart depicting the area allocation within an 8x8 SRAM array with four sets of LUNA-CiM structures. The total area encompassed by the SRAM array and the four LUNA-CiM structures is 3650 $um^2$ square, with each LUNA-CiM structure occupying an area of 287 $um^2$ square. This pie chart effectively visualizes the distribution of area and emphasizes the portion specifically attributed to the added LUNA-CiM units. The chart reveals that the incorporation of these LUNA-CiM structures results in an area overhead comprising 32 percent of the total area of the SRAM array.

\section{Conclusion}

In conclusion, this paper presents a pioneering effort in the development of a design methodology for LUNA-CiM, which offers inherent advantages such as fast and programmable in-memory processing, large noise margin, and compatibility with existing digital design flows and standard-cell libraries. The proposed framework incorporates innovative components that enable low-overhead programmable CiM. By utilizing traditional SRAM cells within the array, LUNA-CiM provides an adaptable storage solution for both data and LUT functions. This approach offers a fully programmable solution for neural acceleration architectures, leveraging the benefits of standard memory infrastructure. Furthermore, the extension of the concept of malleable memory enhances the scalability of LUT-based computation. Through a divide-and-conquer strategy, high-precision multiplications are efficiently transformed into combinations of lower-precision operations, optimizing the overall performance and resource utilization. Experimental evaluations demonstrate the effectiveness of the presented techniques. When compared to traditional LUT-based methods, the proposed approaches yield significant improvements in terms of area overheads and energy consumption. This research represents a significant advancement in the field of efficient and scalable neural processing architectures within SRAM arrays for in-memory computing. The presented LUNA-CiM design methodology offers a promising solution for programmable CiM applications, addressing key challenges such as energy efficiency, scalability, and compatibility with existing design frameworks. Future work can explore further optimizations and applications of LUNA-CiM, paving the way for enhanced neural processing capabilities in various computing domains.
\bibliographystyle{IEEEtran}
\bibliography{main}
\end{document}